\documentstyle{mn}
\def\spose#1{\hbox to 0pt{#1\hss}}
\def\approxlt{\mathrel{\spose{\lower 3pt\hbox{$\sim$}}
        \raise 2.0pt\hbox{$<$}}}
\def\approxgt{\mathrel{\spose{\lower 3pt\hbox{$\sim$}}
        \raise 2.0pt\hbox{$>$}}}
\def\refitem{\par\parskip 0pt\noindent\hangindent 20pt}

\title[Limits from rapid TeV variability of Mrk 421] 
{Limits from rapid TeV variability of Mrk 421}

\author[A. Celotti, A.C. Fabian, M.J. Rees]
{A. Celotti$^1$\thanks{E-mail: {\tt celotti@sissa.it}},
A.C. Fabian$^2$ and M.J. Rees$^2$\\
$^1$S.I.S.S.A., via Beirut 2--4, 34014 Trieste, Italy \\
$^2$Institute of Astronomy, Madingley Road, Cambridge CB3 0HA \\}

\begin{document} 
\maketitle 

\begin{abstract} The extreme variability event in the TeV emission of Mrk
421, recently reported by the Whipple team, imposes the tightest limits on
the typical size of the TeV emitting regions in Active Galactic Nuclei
(AGN). We examine the consequences that this imposes on the bulk Lorentz
factor of the emitting plasma and on the radiation fields present in the
central region of this Active Nucleus.  No strong evidence is found for
extreme Lorentz factors. However, energetics arguments suggest that any
accretion in Mrk 421 has to take place at small rates, compatible with an
advection--dominated regime.  \end{abstract}


\section{Introduction} The intense $\gamma$--ray emission from radio--loud
AGN (and in particular from blazars) argues strongly,
independent of other evidence, in favour of the role of relativistic
beaming in the physics of this class of source, as proposed almost 20
years ago by Blandford \& Rees (1978).  In fact, the requirement that the
$\gamma$--ray source is optically thin to the process of
electron--positron (e$^{\pm}$) pair production, sets strong limits on the
comoving radiation density at frequencies above the pair production
threshold. 

The (potential) target photons can be modelled within the $\gamma$--ray
emitting region, as well as at all scales along the line of sight, from
the very central core of these Active Galaxies to the diffuse background
field. 

Here we concentrate on the physical constraints that these observations
impose within the inner $\sim$ pc (for considerations relative to the
background radiation field see e.g. De Jager, Stecker \& Salamon 1994). In
particular, we discuss the implications of the recent observation of the
large and unprecedented fast variability event in the TeV emission of the
BL Lac object Mrk 421 (z=0.031) by the Whipple Observatory (Gaidos et al. 
1996). On 1996 May 15 the flux above 350 GeV varied by a factor of 20--25
in about 20--30 minutes, with a doubling timescale $< 15$ min. Note that
the whole flare has been observed. This variation sets tight upper limits
on the size of the TeV emitting region. 

\section{Pair opacity constraints}

The optical depth to e$^{\pm}$ pair production by photon--photon
interaction, $\tau_{\gamma-\gamma}$, for $\gamma$--ray photons of energy
$x$ (in units of the electron rest mass energy) is proportional to the
compactness of target photons, $\propto L/R$, where $R$ is the length of
the path of the $\gamma$--ray in the target photon field.  More precisely
\begin{equation}\tau_{\gamma\gamma}(x)\simeq \eta(\alpha) \sigma_T n(x_*) 
x_* R \end{equation} where $x_*$ is the energy (in units of $m_{\rm
e}c^2$) of target photons with number density $n(x_*)$, $\sigma_{\rm T}$
is the Thomson scattering cross section and $\eta(1)\sim 0.12$, where
hereafter an energy spectral index $\alpha\sim 1$ is assumed (Svensson
1987).

\subsection{Radiation field internal to the $\gamma$--ray emitting region}

Let us consider the limits deriving from any radiation field present in
the same region where the TeV photons are produced.

An upper limit to the region size $r_{\gamma}$, implied by the observed
$\gamma$--ray variability, is $R= r_{\gamma}\approxlt c t_{\rm var} \delta
(1+z)^{-1}$, where $\delta\equiv (\Gamma-\sqrt{\Gamma^2-1}
\cos\theta)^{-1}$ is the relativistic Doppler factor of the emitting plasma
moving with a Lorentz factor $\Gamma$ at an angle $\theta$ with respect to
the line of sight and $t_{\rm var}$ is the observed variability timescale
\footnote{Note that the size $r_{\gamma}$ is estimated from the
$\gamma$--ray variability and does not need to assume co-spatial emission
with variable flux at other frequencies.}.  We consider the target photons
at the threshold energy for pair production, where the cross section is
maximum.  Taking into account the Doppler transformations of the observed
$\gamma$--rays, the corresponding target photon frequency is $\nu_{\rm
ir}\sim 1.2\times 10^{14}\ \nu_{\rm TeV}^{-1}\ \delta^2$ Hz, with $\nu_{\rm
TeV} \simeq 1.2\times 10^{26}$ Hz (corresponding to an energy of 0.5 TeV).
The condition of transparency for the TeV $\gamma$--rays,
$\tau_{\gamma\gamma} \approxlt 1$, then translates into a limit on the
radiation field at $\nu_{\rm ir}$.  By assuming that all the observed
monochromatic flux $F_{\rm obs}(\nu_{\rm ir})$ (erg s$^{-1}$ cm$^{-2}$
Hz$^{-1}$) is indeed produced co--spatially with the $\gamma-$rays, the
constraint imposes a lower limit on the relativistic Doppler amplification
(e.g. Dondi \& Ghisellini 1995; see also Gaidos et al. 1996), namely
\begin{equation} \delta \approxgt 10\, F_{\rm obs}^{1/6}(\nu_{\rm ir})\,
t_{\rm var,3}^{-1/6} \end{equation} where the observed values for $t_{\rm
var} = 1000\ t_{\rm var,3}$ s and $\nu_{\rm ir} F_{\rm obs}(\nu_{\rm
ir})\simeq 4\times 10^{-11}$ erg s$^{-1}$ cm$^{-2}$ have been substituted. 
These correspond to the quiescent state reported by Macomb et al. (1995) 
(at $\sim 1.4\times 10^{14}$ Hz). 

A Doppler factor $\delta\simeq \Gamma\approxgt 10$ would therefore be
required to account for the observed TeV emission if the electron/positron
population emitting at TeV energies also emits the observed IR flux. 
Simultaneous monitoring would be required to establish that (see the
results of the 1994 and 1995 multifrequency campaigns, Macomb et al. 1995,
Buckley et al. 1996).

\subsection{External radiation fields}

Analogously, one can consider constraints both on and from any
(quasi--isotropic) radiation field external to the $\gamma$--ray emitting
region, whose photons can efficiently interact with the TeV radiation.

To estimate the photon--photon optical depth in this case it is necessary
to estimate the compactness of this putative field, over a scale size
which corresponds to the distance (from the central compact object) at
which the high energy emission takes place.  A typical perturbation/shock
would naturally form, following a change e.g. in the bulk Lorentz factor
$\Gamma$ of the plasma in the jet, at a distance which depends on
the intensity and time profile of the velocity change. For a variation
$\Delta\Gamma \sim \Gamma$ occurring over the typical timescale associated
with the dimension $r_{\rm s}$ of the compact object where the jet
presumably forms ($r_{\rm s}$ is the black hole Schwarzschild radius), a
shock would develop at a distance $R= R_{\gamma}\sim r_{\rm s} \Gamma^2
(1+z)^{-1}\, $ \footnote{Here for simplicity it is assumed that the
Lorentz factors of the bulk flow and of the front of any perturbation
coincide, as we do not expect that the two values of $\Gamma$ could differ
significantly - for the purpose of this work - unless the shock front
moves at extremely relativistic speeds relative to the mean flow.}.  Note
that there is a minimum characteristic timescale for AGN variability,
which is given by the light crossing time of the compact object $\sim
r_{\rm s}/c$, and if we assume the observed $\gamma$--ray variability does
indeed correspond to this minimum timescale, than the $\gamma$--ray
emitting region would be located at $R_{\gamma}\sim c\ t_{\rm var}
\Gamma^2 (1+z)^{-1}$. In other words, if the base of the jet varies on
$\sim r_{\rm s}/c$, so would the observed $\gamma$--rays, though
the
scale they come from is $\sim \Gamma^2 r_{\rm s}$. 

\subsubsection{Infrared field}

If one simply assumes that a diffuse IR radiation field is present over
scale $R_{\gamma}$, with a total flux of intensity equal to the
observed $F_{\rm obs}(\nu_{\rm ir})$, then the transparency condition 
imposes again a lower limit on the Lorentz factor (which determines
$R_{\gamma}$ and therefore the compactness of target photons). This limit
corresponds to an extremely high \begin{equation} \Gamma \approxgt
1.2\times 10^3 F_{\rm obs}^{1/2} (\nu_{\rm ir})\, t_{\rm var,3}^{-1/2}
\end{equation} which in turn implies that the TeV emission is produced at
distances of the order of $R_{\gamma} \sim 4\times 10^{19} F_{\rm
obs}(\nu_{\rm ir})\ t_{\rm var,3}^{-1}$ cm.

While there is no obvious reason why the plasma cannot propagate with this
extreme $\Gamma$, the above limit depends on the hypothesis that a photon
flux comparable with the observed one pervades the region.  One can then
reverse the argument, assuming a more `traditional' value of the Lorentz
factor, say $\Gamma\sim 10\ \Gamma_1$ (e.g. Ghisellini et al. 1993), and
deducing an upper limit on the diffuse IR field.  This requires that the
external field, $F_{\rm ext}(\nu_{\rm ir})$, \begin{equation}
\left(\frac{R_{\gamma}}{R_{\rm ir}}\right)F_{\rm ext}(\nu_{\rm
ir})\approxlt 7\times 10^{-5} F_{\rm obs}(\nu_{\rm ir})  \end{equation}
which significantly constraints the existence and properties of an IR
emitter, e.g. in the form of a dusty torus, in the surrounding of the
central object \footnote{In a recent paper Protheroe \& Biermann (1997)
examine in details the effect of pair opacity above a typical AGN torus.}.
In terms of luminosity, this corresponds to $\nu_{\rm ir} L_{\rm
ext}(\nu_{\rm ir}) \approxlt 10^{40}$ erg s$^{-1}$, if the IR emitting
region is located at distances $\approxgt R_{\gamma}\sim 3\times 10^{15}
t_{\rm var,3} \Gamma_{1}^2$ cm.

\subsubsection{Starlight}

Within the inner $\sim$ pc of the active galaxy, starlight constitutes a
further radiation field in the surroundings of the compact object. While
it is not possible to exactly quantify its intensity, an order of
magnitude estimate, based on the centre of our galaxy, suggests a
starlight radiation density corresponding to a luminosity $L_{\rm
star}\sim 10^{40}$erg s$^{-1}$ within $R_{\rm star}\sim$ 1 pc, with a
spectrum peaking, say, at few$\times 10^{14}$ Hz. This radiation however
does not contribute significantly to the TeV $\gamma$--ray opacity, giving
rise to an optical depth of only $\tau_{\gamma-\gamma}^{\rm star}\sim
10^{-3} L_{\rm star,40} R^{-1}_{\rm star, pc}$.

\subsubsection{Line emission} 

Variable emission in weak broad lines have been detected in the spectra of
some BL Lac objects (Stickel, Fried \& K\"uhr 1993;  Vermeulen et al. 
1995; Robinson \& Corbett 1996). In particular, Morganti, Ulrich \&
Tadhunter (1992) report the detection of a broad emission feature in Mrk
421, with a luminosity (in H$\alpha$ and NII) of $L_{\rm H\alpha+NII}\sim
1.5\times 10^{40}$ erg s$^{-1}$. While during the observation reported by
Morganti et al. a {\it broad} H$\alpha$ line is not clearly resolved, the
detection of this component in the spectra of other BL Lacs would support
the hypothesis that in Mrk 421 the reprocessed emission also is produced
by high velocity gas, supposedly in a broad line region. By adopting, in
analogy with radio--loud quasars (and in fact quite arbitrarily for BL
Lacs), a covering factor of $\sim$ 10 per cent and a Thomson optical depth
$\tau_{\rm T}$ of a scattering diffuse medium of the same order, one can
estimate the corresponding (isotropic) intensity of a ionizing radiation
field localized within a typical $R_{\rm BLR}$. The compactness of this
putative broad line emitting region translates again into a $\gamma$--ray
optical depth $\tau_{\gamma-\gamma}^{\rm line}\sim 2 L_{\rm ion, 41}
R^{-1}_{\rm BLR, 16}$. Also this limit is too weak to set strong
constraints either on the ionizing radiation field or the bulk speed of
the emitting TeV plasma.

It is worth noticing that the estimate of the photoionizing flux
responsible for an observed broad H$\alpha$ line, requires a luminosity of
the order of $10^{40}$ erg s$^{-1}$, not far from the limit derived in
eq.~(4).

\subsubsection{Disc emission} 

There is little evidence of the presence of thermal emission contributing
to the observed radiation in BL Lac objects, probably the strongest
indication being given by the narrow (and possibly broad)  line emission. 
In particular, the limit inferred in eq.~(4), under the assumption that
the plasma bulk Lorentz factor is $\Gamma \approxlt 10$, sets constraints
on the emission from any material accreting, in the form of a disc, onto
the central black hole. These can be translated into limits on the
fundamental parameters regulating the structure and radiative properties
of an accretion disc, namely the black hole mass $M$, accretion rate $\dot
M$ and viscosity.

In view of the numerical complication involved in the self--consistent
computation of the emission at IR frequencies for the full range of the
parameter space of $M, \dot M$, viscosity (e.g. Szuszkiewicz, Malkan \&
Abramowicz 1996 and references therein), we adopt an extremely simplified
approach. The emission from the disc can be roughly described (in $\nu
L(\nu)$) by a power law (with the standard spectral index $\sim 1.3$),
peaking at the energy roughly corresponding to the maximum disc
temperature, namely between the optical--UV and the soft--X band and
extending to the low energies where the emission from the external radii
dominates.

We therefore consider the limits in mass and accretion rate of a disc not
exceeding the luminosity of Mrk 421 in the optical--soft X--ray band,
$\nu_{\rm uv} L_{\rm uv}$, and extrapolate the emission from the peak
frequency to the IR band as a power law of index 1.3.  This extrapolated
luminosity is then compared with the limit given by eq.~(4).

However it is also necessary to estimate the location (within the limits
of the optically thick Keplerian $\alpha$ disc assumptions, Shakura \&
Sunyaev 1973), of the IR emitting region, $R_{\rm ir}$. In fact, that
allows us: a) to determine self-consistently the solution of the disk
equations; b) to check that the IR emitting region is within the outer
self--gravitating radius; c) most important, to estimate its location with
respect to the $\gamma$--ray emitting region. The last point in fact
determines the interaction angles (and therefore energies) of the target
photons for the TeV $\gamma$--rays. In particular, for $R_{\rm ir} <
R_{\gamma}$, we consider that the absorbed IR photons have been mostly
scattered and isotropized by an external medium with Thomson optical depth
$\tau_{\rm T}\sim 0.1\tau_{\rm T,-1}$. This in turn implies that the IR
disc luminosity, consistent with the observed TeV emission, can be up to
$\sim$ 10 times the limit derived in eq.~(4). 

Let us express the mass $m$ in solar units and the accretion rate in
Eddington units, $\dot m \equiv \dot M/\dot M_{\rm E}$ (where $\dot M_{\rm
E}= L_{\rm E} c^{-2}$ and $L_{\rm E}$ is the Eddington luminosity). If we
require that this bolometric luminosity does not exceed the observed
emission $\nu_{\rm uv} L(\nu_{\rm uv})$, this implies $\dot m\
m_7\approxlt 8\times 10^{-2}$. By extrapolating from the peak of the
energy distribution to IR frequencies, the parameters of a standard thin
Keplerian disc are then constrained to $\dot m\ m_7^2 \approxlt 5\times
10^{-3} \tau_{\rm T,-1}^{-3/2}$. In this limit we have taken
into account that for typical $m\approxgt 10^6$ and $\dot m \approxlt 1$,
the disc radius which dominates the IR emission is smaller that
$R_{\gamma}$ (where a Lorentz factor $\Gamma=10$ has been adopted). For
this range of $m$, $\dot m$, and a viscosity parameter $\approxgt
10^{-5}$, the disc at $R_{\rm ir}$ is not self--gravitating.

Therefore unless either Mrk 421 harbours a black hole of mass $\approxlt
10^6 M_{\odot}$ or the Lorentz factor of the $\gamma$--ray emitting plasma
is $\Gamma\gg 10$ or the TeV $\gamma$--ray emitting region is at much
larger distances, we are forced to conclude that the accretion rate 
in
this object has $\dot m \approxlt 10^{-2}-10^{-3}$.

On the other hand, observations on VLBI scales lead to estimates of the
power emitted in the form of jet kinetic luminosity exceeding 
$\sim 10^{46}$ erg s$^{-1}$ (e.g. Celotti, Padovani \& Ghisellini
1997), suggesting that indeed Mrk 421 is harboring a much higher mass
object. This could be reconciled with the low radiative (quasi--thermal) 
power if any accreting disc is radiatively inefficient. The deduced limits
on $\dot m$ are then consistent with the accretion occurring in the
advection--dominated regime.

\subsubsection{Conclusions} 

The constraints derived in this section imply that if a region comparable
in size to the TeV production site $R_{\gamma}$ is pervaded by a radiation
field of intensity comparable with the observed one, then an extremely
high bulk Lorentz factor, $\Gamma\approxgt 10^3$, is required in order to
overcome the limits from photon--photon opacity.

However, one can easily envisage alternative possibilities which would not
set such strong limits on the relativistic motion of the plasma. Namely: 

\noindent a) any quasi--isotropic IR field could be produced on scales
much larger than $R_{\gamma}$. $\Gamma\sim 10$ would require $R_{\rm
ir}\sim 10^4 R_{\gamma}$ ($\sim 10^{19}\ \Gamma^2_{10}\ t_{\rm var,3}$
cm).

\noindent b) any isotropic (non--beamed) component is much weaker than the
observed luminosity, and a typical $\Gamma\sim 10$ requires that $ F_{\rm
ext}(\nu_{\rm ir}) \approxlt 10^{-4} F_{\rm obs}(\nu_{\rm ir})$. This
result implies that any disc in the nucleus of Mrk 421 accretes at a rate
$\dot m\approxlt 10^{-3}$ for black hole masses $M\approxgt 10^7
M_{\odot}$. 

\noindent Finally, radiation of stellar origin and a possible
photoionizing continuum do not significantly contribute to the opacity for
TeV $\gamma$--rays. 

\section{On the TeV emission mechanism}

Let us now briefly consider the implications of the observed variability
on the radiation mechanism producing the TeV emission. It is widely (but
not universally) believed that the high energy $\gamma$--ray component in
the spectra of blazars originates from inverse Compton scattering of
relativistic electrons/positrons on soft photons (e.g. Sikora 1994 for a
review). The origin of this photon field however is still a matter of
debate, it could comprise both synchrotron photons (SSC) and an isotropic
photon field external to the synchrotron emitting region. 

The condition that, in the comoving frame, seed photons of frequency
$\nu_o$ are efficiently scattered to TeV energies in the Thomson regime,
implies that $\Gamma \gamma \approxgt 2\times 10^6$ and $\nu_o \approxlt
2\times 10^{13}$ Hz, where $\gamma m_{\rm e} c^2$ is the energy of the
scattering electrons/positrons in the jet frame. 

The energy density of photons at $\nu_o$, as seen by the emitting plasma,
can be estimated as $U'_{\rm ext} \sim 3\times 10^{-2} L_{\rm ext,40}
\tau_{\rm T, -1} \Gamma_{1}^{-2} t_{\rm var,3}^{-2}$ erg cm$^{-3}$ where a
typical $\Gamma\sim 10 \Gamma_{1}$ has been assumed and the dimension
corresponding to the distance of the $\gamma$--ray emitting region. From
the observed luminosity, $L_{\rm obs}\sim 2\times 10^{44} L_{\rm obs,44}$
erg s$^{-1}$, the synchrotron photon energy density is of the order of $
U'_{\rm syn} \sim 0.6\ L_{\rm obs,44}\ t_{\rm var,3}^{-2}\
\delta_{10}^{-6}$ erg cm$^{-3}$. Therefore $U'_{\rm syn}/U'_{\rm ext}\sim
20\ \delta_{10}^{-4}$ which implies that the scattering of external
photons can dominate the SSC emission only for $\delta \approxgt 20$.

Finally, rough equilibrium of the total luminosities emitted by Mrk 421 in
the synchrotron and inverse Compton components of the spectrum (peaking in
the UV--soft X--ray and $\gamma$--ray bands, respectively) implies a
similarity of the magnetic and radiation energy densities, if the two
radiative components are emitted in the same spatial region and most of
the inverse Compton scattering is in the Thomson regime. This requires
typical magnetic fields of the order of a few Gauss.
 
\section{Summary and discussion}

Consequences of the extraordinary variability event detected by Whipple
for the physical properties of the emitting plasma and the radiation fields
which surround the compact object and the powering of the central black
hole in Mrk 421 have been examined. 

We find that high ($\approxgt 10$) or even extreme ($\approxgt 10^3$) 
bulk Lorentz factors are required for the source to be transparent to TeV
$\gamma$--rays, unless the radiation fields within and around the
$\gamma$--ray emitting region are much lower than inferred from the
observed flux.

It should be recalled that BL Lac objects in general show little evidence
of strong diffuse/isotropic radiation, and therefore the above
considerations are not (yet) evidence for the most extreme values of the
Lorentz factor. However we note that due to the weak dependence of the
Lorentz factor on the compactness in target photons, we do not expect that
such high $\Gamma$ can be easily derived, at present, by $\gamma$--ray
variability measures. 

Still there are no reasons a priori to exclude such a possibility. 
Clearly, the beaming angle corresponding to $\Gamma$ as high as thousands
would create serious statistical inconsistencies if these results would be
extended to all sources and the effective opening angle of the jet was of
the same order. Indeed it is likely that the jet plasma flows within an
angle $\gg 10^{-3} \Gamma_{3}^{-1}$. Furthermore, it is reasonable to
imagine that jets reach such speeds only in the very inner core and that,
in Mrk 421, we are observing the jet virtually aligned with its axis. This
possibility would be also consistent with the mild apparent superluminal
velocity $\beta_{\rm app} c\sim 3.8\ c$ observed in the radio components
of this source (Zhang \& Baath 1990). 

One should note that even the weaker of the constraints derived on
$\Gamma$ is significantly higher than typical Lorentz factors in BL Lac
objects, as derived from limits on the SSC flux, VLBI radio measurement
and statistical arguments (e.g.  Ghisellini et al. 1993). Furthermore for
these $\Gamma$ the TeV emission is likely to be mainly due to SSC. 

Finally, we briefly consider the global energetics.  We have shown
that the assumption of a more `traditional' value of the bulk Lorentz
factor implies such stringent limits on the radiation field surrounding
the nucleus that any accretion is likely to occur at rates $\dot
m\approxlt 10^{-3}$. Interestingly this is further supported by
independent recent findings. In fact, following the ideas proposed by Rees
et al. (1982) and Fabian \& Rees (1995), it has been suggested (Reynolds
et al. 1996a) that M87, the nearby FR~I galaxy, and by extension FR~I
radio galaxies in general, host an active nucleus of very low radiative
efficiency as a consequence of the accretion being advection--dominated. 
Extrapolating this to the plausible beamed counterparts of FR~I galaxies,
namely BL Lac objects (e.g. Urry \& Padovani 1995 for a review), one
consequently predicts that these sources also are powered by an
advection--dominated flow which is responsible for their low radiative
(disk) efficiency. The jet power could then be extracted, analogously to
M87, either from the spin energy of the black hole (Blandford \& Znajek
1977) or as Poynting--dominated outflow along the system rotation axis of
an accretion--powered object (Blandford \& Payne 1982). This power could
be of the order of $L_{\rm em} \sim 10^{43} (a/m_g)^2 B_2^2 M_9^2$, where
$B_2\propto \dot m_{-3}^{1/2}$ is the equipartition field (e.g.  Narayan
\& Yi 1995), $\dot m=10^{-3} \dot m_{-3}$, $ac$ is the specific angular
momentum and $m_g=GM/c^2$ the gravitational radius. \footnote{Note that
the modeling of the high energy spectral distribution of BL Lacs favors a
rather low magnetic field (in the emitting region) compared to high power
sources.}

Indeed we note that BL Lacs, and in particular Mrk 421, are hosted in
elliptical galaxies, consistently with high values of the black hole mass
and the suggestion by the above authors that these systems could be
accreting hot interstellar medium quasi--spherically at the (Bondi)
accretion rate.

This hypothesis could also be consistent with the lack of much
reprocessing gas in their surrounding as well as a low ionization flux
(but note earlier comments on broad lines, Sec. 2.2.3). If BL Lacs
correspond to the final evolutionary stage of sources accreting through a
radiatively efficient geometrically thin disk, this would be in
(qualitative) agreement with the redshift distribution of BL Lacs, which
is biased to low redshifts.

The jet power deduced above could be compared with that estimated from the
density of emitting electrons and the inferred Lorentz factors on VLBI
scales, which imply a jet power of about three orders of magnitude higher,
strongly suggesting that an high mass black hole lies in the very central
nucleus of Mrk 421.  The high value of the kinetic jet power can however
also be reconciled with luminosities $\sim 10^{43}$ erg s$^{-1}$ if jets
of BL Lacs (and FR~I) are mainly composed of an e$^{\pm}$ plasma, as
recently suggested on other grounds (Celotti et al. 1997, Bodo et al. in
preparation) and in particular for M87 (Reynolds et al. 1996b).

As it is often the case, some of the possibilities discussed above could
be settled when more and higher quality data become available. In
particular we emphasize the importance of observations of variability at
very high energies, which lead to much stronger physical implications when
performed simultaneously with other energy bands.

\section*{Acknowledgments} 

We thank M. Hillas for discussing with us the results of Whipple
observations and Mitch Begelman for reading the manuscript. The Italian
MURST (AC) and the Royal Society (ACF and MJR) are acknowledged for
financial support. 

\section*{References} 

\refitem Blandford R.D., Payne D.G., 1982, MNRAS, 199, 883

\refitem Blandford R.D., Rees M.J., 1978, in A. M. Wolfe, ed,
Proc. Pittsburgh Conf. on BL Lac Objects. Univ. of Pittsburgh Press,
Pittsburgh, p. 328

\refitem Blandford R.D., Znajek R.L., 1977, 179, 433

\refitem Buckley J.H., et al., 1996, ApJ, 472, L9

\refitem Celotti A., Padovani P., Ghisellini G., 1997, MNRAS, in press

\refitem De Jager O.C., Stecker F.W., Salamon M.H., Nat, 1994, 369, 294

\refitem Dondi L., Ghisellini G., 1995, MNRAS, 273, 583

\refitem Fabian A.C., Rees M.J., 1995, MNRAS, 277, L55

\refitem Gaidos J.A., et al., 1996, Nat, 383, 319

\refitem Ghisellini G., Padovani P., Celotti A., Maraschi L., 1993, ApJ,
407, 65

\refitem Macomb D.J., et al., 1995, ApJ, 449, L99 [Erratum: 1996, ApJ, 459, 
L111]

\refitem Morganti R., Ulrich M.-H., Tadhunter C.N., 1992, MNRAS, 254, 546 

\refitem Narayan R., Yi I., 1995, ApJ, 452, 710

\refitem Protheroe R.J., Biermann P.L., 1997, Astroparticle Phys., in press

\refitem Rees M.J., Begelman M.C., Blandford R.D., Phinney E.S., 1982,
Nat, 295, 17

\refitem Reynolds C.S., Fabian A.C., Rees M.J., Celotti A., 1996b,
MNRAS, 283, 873

\refitem Reynolds C.S., Di Matteo T., Fabian A.C., Hwang U., Canizares C.R., 
1996a, MNRAS, 283, 111

\refitem Robinson A., Corbett E., 1996, IAU Circ. 6463

\refitem Shakura N.I., Sunyaev R.A., 1973, A\&A, 24, 337

\refitem Sikora M., 1994, ApJS, 90, 923

\refitem Sikora M., Begelman M.C., Rees, M.J., 1994, ApJ, 421, 153

\refitem Stickel M., Fried J.W., K\"uhr H., 1993, A\&AS, 98, 393

\refitem Svensson R., 1987, MNRAS, 277, 403 

\refitem Szuszkiewicz E., Malkan M.A., Abramowicz M.A., 1996, ApJ, 458, 474

\refitem Urry C.M., Padovani P., 1995, PASP, 107, 803

\refitem Vermeulen R.C., Ogle P.M., Tran H.D., Browne I.W.A., Cohen
M.H., Readhead A.C.S., Taylor G.B., 1995, ApJ, 452, L5

\refitem Zhang F.J., Baath L.B., 1990, A\&A, 236, 47

\end{document}